\pdfoutput=1

\documentclass[aps, pra, preprint, tightenlines, superscriptaddress]{revtex4-1}
\usepackage{amsmath,amssymb,graphicx}

\begin{document}

\title{Silicon-nitride photonic circuits interfaced with monolayer MoS$_{2}$}

\author{Guohua Wei}
\affiliation{Applied Physics Program, Northwestern University, 2145 Sheridan Road, Evanston, IL 60208, USA}
\author{Teodor K. Stanev}
\affiliation{Department of Physics and Astronomy, Northwestern University, 2145 Sheridan Road, Evanston, IL 60208, USA}
\author{David A. Czaplewski}
\affiliation{Center for Nanoscale Materials, Argonne National Laboratory, 9700 S Cass Avenue, Argonne, IL 60439, USA\bigskip\bigskip}
\author{Il Woong Jung}
\affiliation{Center for Nanoscale Materials, Argonne National Laboratory, 9700 S Cass Avenue, Argonne, IL 60439, USA\bigskip\bigskip}
\author{Nathaniel P. Stern}\email{Corresponding author: n-stern@northwestern.edu}
\affiliation{Applied Physics Program, Northwestern University, 2145 Sheridan Road, Evanston, IL 60208, USA}
\affiliation{Department of Physics and Astronomy, Northwestern University, 2145 Sheridan Road, Evanston, IL 60208, USA}

\begin{abstract}
We report on the integration of monolayer molybdenum disulphide with silicon nitride microresonators assembled by visco-elastic layer transfer techniques. Evanescent coupling from the resonator mode to the monolayer is confirmed through measurements of cavity transmission. The absorption of the monolayer semiconductor flakes in this geometry is determined to be 850 dB/cm, which is larger than that of graphene and black phosphorus with the same thickness. This technique can be applied to diverse monolayer semiconductors for assembling hybrid optoelectronic devices such as photodetectors and modulators operating over a wide spectral range.
\end{abstract}

\maketitle 

With traditional silicon transistors approaching fundamental size limits, integrated electro-optical circuits are a promising alternative for low-power, high-speed information processing~\cite{assefa201290nm}. Hybrid electro-optical devices optimized for light modulation and detection can be assembled by integrating atomic-scale materials such as graphene with on-chip photonic circuits~\cite{liu2011graphene,wang2013high,gan2013chip}.  Although the high carrier mobility and gapless band structure of graphene enable high frequency and broadband operation, possible photonic applications are limited by the absence of an appreciable bandgap.  In contrast, other two-dimensional (2D) materials such as single layers of transition metal dichalcogenides~(TMDCs) exhibit direct bandgaps ranging from 1.1~eV to $>$ 2.0~eV ~\cite{wang2012electronics,xia2014two}. This feature allows ultra-high on/off ratio transistors and sensitive photodetectors based on TMDC layers~\cite{radisavljevic2011single,lopez2013ultrasensitive}. The diverse electronic properties of 2D materials can be further exploited for high-performance opto-electronics by assembling layered heterostructures~\cite{geim2013van,wang2015van} or for new polarization-sensitive valley-dependent devices~\cite{xiao2012coupled}. Interfacing TMDCs with optical waveguides and resonators is required to take advantage of these layer-sensitive structures in photonic circuits.

Emission of 2D TMDCs can be controlled by interfacing layers with optical resonators ~\cite{liu2015strong,gan2013controlling}. Lasing from TMDCs coupled with photonic cavities has been recently demonstrated~\cite{wu2015monolayer,ye2015monolayer}, but the coupling of TMDCs to planar photonic circuits has not been explored. Absorptive or refractive integrated TMDC-photonic devices such as modulators or photodetectors require evanescent coupling of optical modes with monolayer materials. Strong absorption can reduce the device footprint and increase modulation depth or sensitivity. Here, we adopt visco-elastic transfer techniques~\cite{castellanos2014deterministic} to assemble an on-chip silicon nitride integrated photonic circuit architecture evanescently coupled to monolayer molybdenum disulphide (MoS$_{2}$).  We measure strong absorption of the monolayer in the evanescent field larger than that for monolayer graphene or black phosphorus. The integration of 2D semiconductors with silicon-based photonic structures demonstrates a path for assembling on-chip functional hybrid electro-optical devices based on monolayer materials and their heterostructures.

\begin{figure}[tb]
\centerline{\includegraphics[width=0.5\columnwidth]{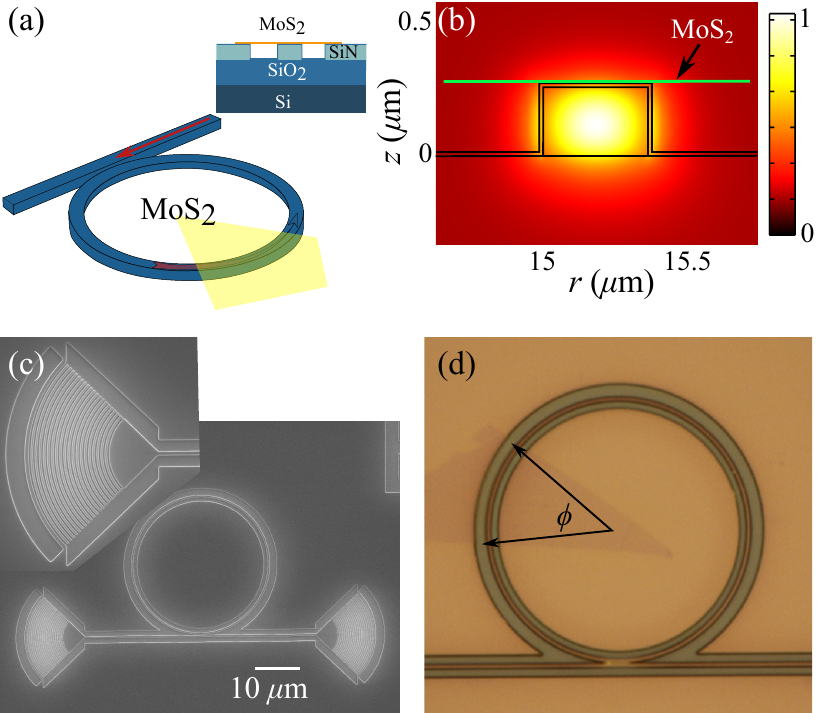}}
\caption{(a)~A schematic of the experimental set--up. Insert shows the device profile. (b)~Mode profile of the fundamental TE-like mode in the ring with MoS$_{2}$ monolayer on top. (c)~SEM image of a representative device. (d)~Optical image of a device with the ring covered by a monolayer MoS$_{2}$ over a circumference corresponding to $\phi = \pi/4$. The ring radius is 15 $\mu$m.}
\label{fig:1}
\end{figure}

Evanescent coupling of nanomaterials in hybrid photonic devices can be confirmed by measuring absorption. Direct measurement of absorption loss from monolayer TMDCs is challenging because of the small overlap of atomic-scale layers with evanescent fields of guided optical modes. As demonstrated with graphene, integrated photonic structures such as Mach--–Zehnder interferometers and microresonators can be exploited for measuring coupling since they are sensitive to single-layer absorption~\cite{gruhler2013high,liu2011graphene,wang2013high}. Monolayer MoS$_2$ is even more attractive for this mode of measurement because of its enhanced band edge absorption of 5-10\% for normal incident light~\cite{bernardi2013}, which is several times larger than the broadband absorption of graphene~\cite{gruhler2013high}. Since the \textit{Q}--factor of a photonic resonator is closely related to the absorption losses in its mode, microring resonators with high $Q$--factors are a sensitive system for measuring coupling of monolayer materials to optical modes~(Fig.~\ref{fig:1}(a)). For demonstration of photonic coupling, we use the canonical monolayer MoS$_2$, although the technical approach directly translates to other TMDCs without modification.

Since the direct bandgaps of various 2D TMDCs are distributed from 1.1 eV (MoTe$_{2}$) to larger than 2.0 eV (WS$_{2}$), a material platform for TMDC hybrid photonics needs low visible and infrared light absorption. Silicon nitride (SiN) is a wide bandgap semiconductor material transparent to both visible and infrared wavelengths. Because of its large refractive index ($n \approx 2.0$) contrast to SiO$_{2}$ ($n \approx 1.45$) and its compatibility with CMOS processing, SiN is very attractive for linear and non-linear photonics in wavelength regimes where traditional silicon photonics architecture would absorb~\cite{moss2013new}. Here, we use stoichiometric Si$_{3}$N$_{4}$ deposited by low pressure chemical vapor deposition on silicon substrates. Because MoS$_{2}$ is observed to have poor adhesion to SiN, a layer of 15~nm PECVD SiO$_{2}$ is deposited over the SiN to improve adhesion of the MoS$_{2}$ after device patterning.

Resonators suitable for measuring material absorption must simultaneously exhibit both well-resolved changes in $Q$ from the layer transfer process and ideally-tuned external coupling for a spectral measurement. These conditions require tailoring the microresonator for the range of expected material parameters. We estimate the expected resonator loss caused by MoS$_2$ with finite element modeling using COMSOL Multiphysics.  The monolayer MoS$_2$ is approximated as a rectangular cross-section of thickness 0.65 nm above the SiO$_2$-capped SiN device (Fig.~\ref{fig:1}(a) inset). The resonator is simulated using an axisymmetric model with a SiN cross-section 375 nm wide and 250 nm high. MoS$_2$ dielectric parameters, specifically the real and imaginary parts of permittivity, were taken from~\cite{li2014measurement}.  The effective absorption loss $\alpha$ of the monolayer is extracted from the complex eigenfrequency of the simulated resonance mode~(Fig.~\ref{fig:1}(b)). The simulated material absorption is $\alpha = 1390$ dB/cm at a wavelength near 660~nm with the ring radius of $R = 15$ $\mu$m. Based on the transmission spectra expected before and after monolayer transfer, ring radii of 12 and 15~$\mu$m were chosen for fabrication. Rings resonators of this geometry have negligible bending loss, suitable external waveguide mode coupling, and a free spectral range commensurate with our measurement apparatus.

Ring resonators of the desired geometry are patterned with high-resolution electron beam lithography using a JEOL~9300 operating at 100 KV with a shot pitch of 3~nm~(Fig.~\ref{fig:1}(c)). Positive resist (GL2000) was used for pattern definition and as the etching mask following development. This writing procedure creates channels on both sides of the waveguides that the transferred monolayer bridges~(Fig.~\ref{fig:1}). The ring resonators are coupled to planar bus waveguides with the same cross-sectional dimensions as the ring (375 nm $\times$ 250 nm). A pair of grating couplers terminate the two ends of the bus waveguide for input and output coupling. The device is carefully designed so both can be in the field of view of a 100$\times$ objective.

Visco-elastic transfer is employed for deterministic assembly of monolayer integrated photonic devices~\cite{castellanos2014deterministic}. Monolayers of MoS$_{2}$ are mechanically exfoliated from a bulk crystal with a gel film and positioned to the desired location using a 3-axis translation stage. The gel is lowered into contact with the substrate and slowly peeled away, leaving the monolayer flake on top of the ring resonator as confirmed by optical microscopy. Monolayers can be placed with micrometer accuracy on top of photonic structures. Fig.~\ref{fig:1}(d) shows a representative device covered by a monolayer of MoS$_{2}$ with $\phi~=~\pi/4$. The MoS$_{2}$ layer is suspended over the trenches surrounding the ring. The nearby SiN on either side of the trench helps support the intact monolayer so that it remains wrinkle-free with minimal strain near the coupling region.

\begin{figure}[tbp]
\centerline{\includegraphics[width=0.5\columnwidth]{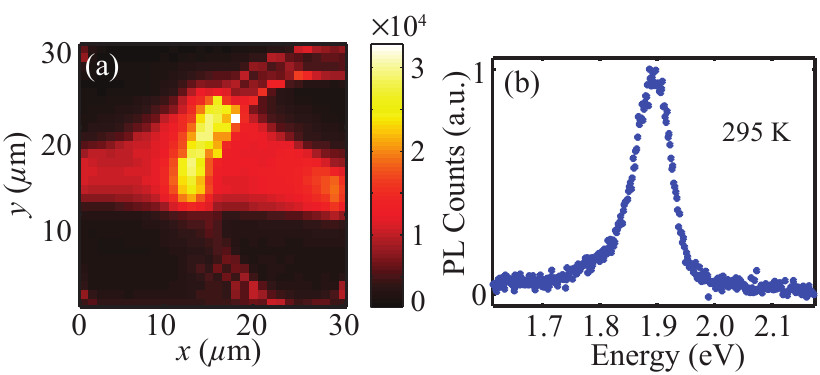}}
\caption{(a)~Photoluminescence intensity map using a 40 $\mu$W pump laser and analyzed with a spectrometer and CCD. The spectra are integrated between 600 nm and 720 nm. (b)~A representative PL spectrum of MoS$_2$ on the ring.}
\label{fig:PL}
\end{figure}

Photoluminescence (PL) characterization is used to confirm successful monolayer transfer. Fig.~\ref{fig:PL}(a) shows spatially-resolved PL from the device shown in Fig.~\ref{fig:1}(d), and a representative spectrum from a single $\sim$ 1 $\mu$m spot on top of the ring is shown in Fig.~\ref{fig:PL}(b). Each spectrum is taken with a scanning microscope with 100$\times$ objective and 40 $\mu$W excitation with a 532 nm laser. The single spectrum clearly shows the expected PL of a monolayer MoS$_{2}$, while the PL map confirms the layer coverage. The PL intensity is higher along the ring/channels where the monolayer is suspended~\cite{mak2010atomically}. Our measurements also show that the ring and channels cause strong scattering of the pump laser which generates stronger PL signal.

\begin{figure}[tbp]
\centerline{\includegraphics[width=0.5\columnwidth]{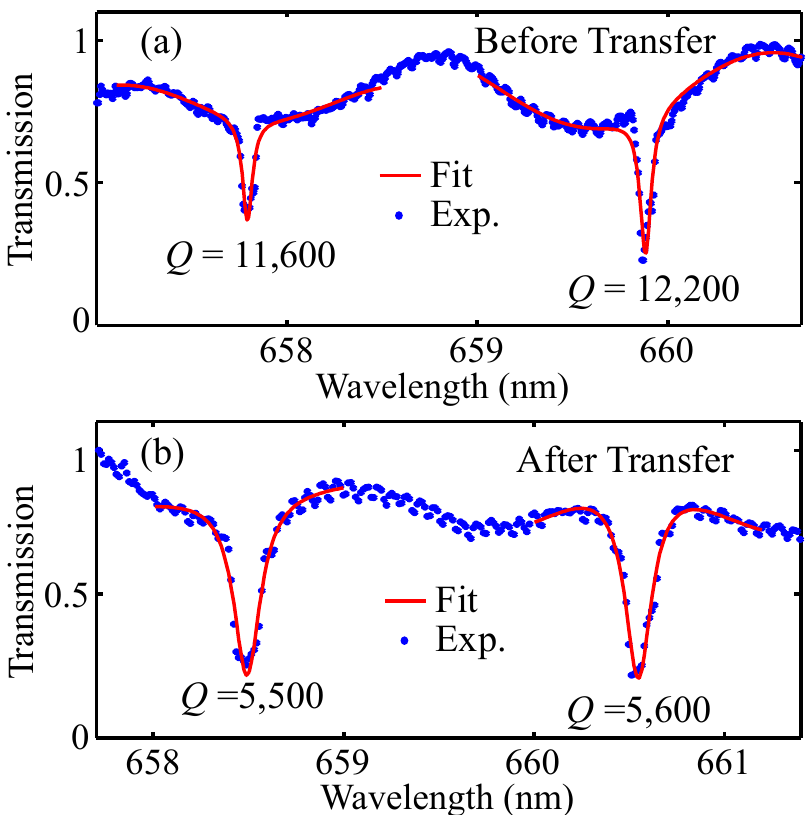}}
\caption{Transmission spectrum of a resonator before~(a) and after~(b) transferring MoS$_{2}$.  Quality factors are extracted from Lorentzian fits to the resonances (red line). }
\label{fig:transmission}
\end{figure}

We fabricated several devices with distributed angle-of-coverage ($\phi$) of monolayer MoS$_2$. Cavity transmission spectra are measured before and after monolayer transfer using a tunable diode laser with linewidth smaller than 1 GHz (which is much narrower than our $>$ 300 GHz resonances linewidths). The output from the grating is spatially filtered for collection. The intrinsic quality factor due only to intracavity absorption and scattering loss is estimated from the transmission spectra assuming no backscattering. Fig.~\ref{fig:transmission} shows a representative transmission spectra of one device with 15$^\circ$ coverage before and after monolayer transfer. We analyze up to two resonant modes for each device, limited by the tuning range of the laser and the free spectral range of the resonators.  Each mode has a somewhat different $Q$--factor. These differences between adjacent resonator modes can be the result of different local mode profiles or wavelength dependence of absorption, which our analysis does not quantitatively account for. The resonance center wavelengths shift after monolayer transfer because the altered mode effective refractive index.

The reduction of the intrinsic \textit{Q}--factor is used to extract the change in absorption loss per length $\varDelta\alpha$ of the ring caused by the monolayer in the evanescent region. The intrinsic $Q$ of an axisymmetric resonator can be expressed in terms of the average loss per unit length $\alpha$:
\begin{equation}
Q=\frac{10}{\textrm{ln(10)}}\frac{2\pi n_{g}}{\lambda_{0}\alpha}
\label{eq:1}
\end{equation}
Here, $\lambda_{0}$ is the resonance wavelength and \textit{n$_{g}$} is the effective refractive index of the guided mode. This internal loss comes from multiple channels such as scattering, bending loss (which is negligible for the size of resonators used), and material absorption loss. Assuming scattering, bending, and SiN loss are essentially unchanged by the layer transfer process, the difference in the effective absorption loss rate before ($\alpha_{i}$) and after ($\alpha_{f}$) monolayer transfer can be written
\begin{equation}
\varDelta\alpha=\alpha_{f}-\alpha_{i}=\frac{2\pi n_{g}}{\lambda_{0}}\frac{10}{\textrm{ln(10)}}\left(\frac{1}{Q_{f}}-\frac{1}{Q_{i}}\right)
\label{eq:2}
\end{equation}
where \textit{Q}$_{i}$ and \textit{Q}$_{f}$ are the \textit{Q}--factors before and after placing monolayer MoS$_{2}$ on the ring. As the ring is only partially covered by MoS$_{2}$, the total effective loss consists of the original total loss around the full ring plus additional absorption loss in the covered region caused by interaction with MoS$_{2}$ ($\alpha_{\textrm{int}}$) :
\begin{equation}
2\pi R\alpha_{f}=2\pi R\alpha_{i}+\phi R\alpha_{\textrm{int}}
\label{eq:3}
\end{equation}
where $\phi$ is the angle by which the monolayer covers the ring resonator and $R$ is the ring radius. We neglect the change in effective mode index and finite mode radial size since the corrections are smaller than the precision of the experiment. The relation between the measured $\varDelta\alpha$ and material absorption is
\begin{eqnarray}
\varDelta\alpha=\alpha_{f}-\alpha_{i}=\alpha_{\textrm{int}}\frac{\phi}{2\pi}
\label{eq:4}
\end{eqnarray}
The change in effective absorption loss rate $\varDelta\alpha$ is proportional to the coverage fraction $\phi/2\pi$ with a coefficient equal to the absorption per unit length of the monolayer  in the evanescent field of this particular geometry.
\begin{figure}[tbp]
\centerline{\includegraphics[]{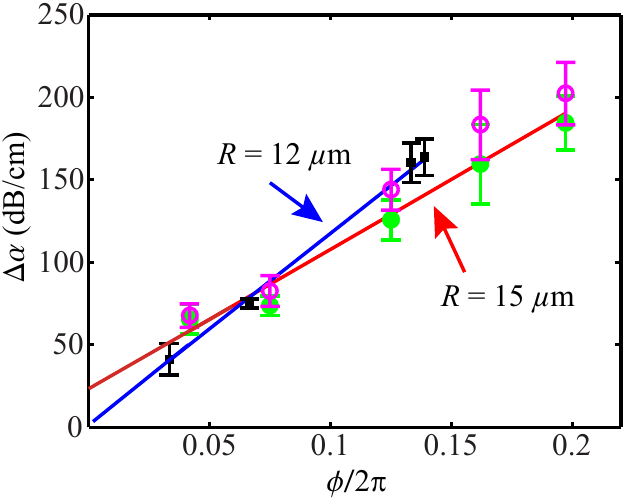}}
\caption{$\varDelta\alpha$ with different coverage angles for different radius of rings. For each $R = 15$ $\mu$m device, the two resonances from each device are differentiated by solid and transparent circles.}
\label{fig:Abs}
\end{figure}

The effective material absorption loss $\varDelta\alpha$ is extracted from the \textit{Q}--factors according to Eq.~\ref{eq:2} (Fig.~\ref{fig:Abs}). A weighted linear fit was performed to each $R$ data set separately, yielding $\alpha_\textrm{int} = 850 \pm 83$~dB/cm ($1157 \pm 48$~dB/cm) with an intercept of $23 \pm 8$~dB/cm ($2 \pm 3$~dB/cm) for resonators with $R = 15$~$\mu$m ($R = 12$~$\mu$m). The non-zero intercept could be caused by scattering due to the discontinuous effective mode index at the boundary between covered and uncovered areas. The difference between $\alpha_\textrm{int}$ for the two radii is statistically significant, but the cause is unknown. The fitted $\alpha_\textrm{int}$ are close to but unambiguously smaller than the $\alpha_\textrm{int}$ = 1390~dB/cm expected from numerical simulations (Fig.~\ref{fig:1}(b)). The discrepancy can have several causes. The monolayers may not be entirely flat as in the simulation or the mode profile may not be exactly as simulated in the axisymmetric model. Absorption and PL of monolayer MoS$_{2}$ are very sensitive to the substrate~\cite{buscema2014effect,sercombe2013optical}. The PL center wavelength on top of our SiN devices is 658~nm, which is higher energy compared to typical results from the literature and from our monolayers on SiO$_2$ substrates~\cite{splendiani2010emerging,mak2010atomically,buscema2014effect,sercombe2013optical}. The absorbance of monolayer MoS$_{2}$ is also sensitive to substrate quality; substrate surface roughness can suppress the absorption or emission of the monolayer~\cite{buscema2014effect}. Our PECVD-deposited capping layer is expected to have high surface roughness~\cite{sercombe2013optical} which can lead to differences in absorption from that predicted in a simulation using literature material parameters.

Despite these quantitative uncertainties, the resonant absorption of monolayer MoS$_2$ in an evanescent field is unambiguously larger than that for monolayer graphene coupled to SiN waveguide circuits (670 dB/cm)~\cite{gruhler2013high}. This result is anticipated because of the increased absorbance of monolayer MoS$_{2}$ (5-10\%) compared to graphene (2-3\%)~\cite{bernardi2013}. The MoS$_2$ absorption is also larger than that measured for waveguide-integrated monolayer black phosphorus~\cite{youngblood2014waveguide}.

In conclusion, we successfully interfaced monolayer MoS$_{2}$ with on-chip photonic circuits through evanescent field coupling. As the absorption of monolayer semiconductors can be modulated by applying a gate voltage~\cite{newaz2013electrical}, the achievement of large evanescent optical coupling between MoS$_2$ and planar photonic resonators suggests that 2D semiconductors are compelling materials for hybrid opto-electronics. Our approach can be applied to combine atomic-scale semiconductors with engineered SiN optical circuits for functional linear or nonlinear photonic devices operating over a wide spectrum and potentially for new classes of polarization-sensitive photonic circuits for modulation and information processing~\cite{lenferink2014coherent}.

This work is supported by the U.S. Department of Energy, Office of Basic Energy Sciences, Division of Materials Sciences and Engineering (DE-SC0012130) (spectroscopy), the Institute for Sustainability and Energy at Northwestern (opto-electronic device integration), and Argonne National Laboratory. Use of the Center for Nanoscale Materials was supported by the U. S. Department of Energy, Office of Science, Office of Basic Energy Sciences, under Contract No. DE-AC02-06CH11357. This work utilized Northwestern University Micro/Nano Fabrication Facility (NUFAB), which is supported by the State of Illinois and Northwestern University.  N.P.S. is an Alfred P. Sloan Research Fellow.

\end{document}